\pdfoutput=1

\documentclass[11pt,twoside,a4paper,cmspaper,final,collab]{cms-tdr}
\def\svnVersion{57407:57425M}\def\cmsCernNo{2011-074}\def\cmsCernDate{\today}\def\cmsMessage{Submitted to Physical Review Letters}

\begin{document}\cmsNoteHeader{HIN-11-007}

\hyphenation{had-ron-i-za-tion}
\hyphenation{cal-or-i-me-ter}
\hyphenation{de-vices}

\RCS$Revision: 57425 $
\RCS$HeadURL: svn+ssh://alverson@svn.cern.ch/reps/tdr2/papers/HIN-11-007/trunk/HIN-11-007.tex $
\RCS$Id: HIN-11-007.tex 57425 2011-05-24 19:18:57Z alverson $
\cmsNoteHeader{HIN-11-007} 
\title{\texorpdfstring{Suppression of excited $\Upsilon$ states in PbPb collisions at $\sqrt{s_{\rm NN}} = 2.76$~TeV}{Suppression of Upsilon excited states in PbPb collisions at a nucleon-nucleon centre-of-mass energy of 2.76 TeV}}

\author[cern]{The CMS Collaboration}

\date{\today}

\abstract{
A comparison of the relative yields of $\Upsilon$ resonances in the $\mu^+\mu^-$ decay channel in PbPb and \Pp\Pp\ collisions at a centre-of-mass energy per nucleon pair of 2.76~TeV, is performed with data collected with the CMS detector at the LHC. Using muons of transverse momentum above 4~GeV/$c$ and pseudorapidity below 2.4, the double ratio of the $\Upsilon(2S)$ and $\Upsilon(3S)$ excited states to the $\Upsilon(1S)$ ground state in PbPb and \Pp\Pp\ collisions, $[\Upsilon(2S+3S)/\Upsilon(1S)]_{\rm PbPb} / [\Upsilon(2S+3S)/\Upsilon(1S)]_{\Pp\Pp}$, is found to be 0.31~$_{-0.15}^{+0.19}$~(stat.)~$\pm$~0.03~(syst.). The probability to obtain the measured value, or lower, if the true double ratio is unity, is calculated to be less than 1\%.
}

\hypersetup{%
pdfauthor={CMS Collaboration},%
pdftitle={Suppression of Upsilon excited states in PbPb collisions at a nucleon-nucleon centre-of-mass energy of 2.76 TeV},%
pdfsubject={CMS},%
pdfkeywords={CMS, physics, upsilon, heavy ions}}

\maketitle 

\hyphenation{ana-ly-sis}
\hyphenation{ana-ly-ses}
\hyphenation{de-tec-tors}
\hyphenation{re-so-lu-tion}
\hyphenation{re-so-nan-ces}

Quantum chromodynamics (QCD) predicts that strongly interacting matter undergoes a phase transition to a deconfined state, often referred to as the quark-gluon plasma (QGP), in which quarks and gluons are no longer bound within hadrons. Calculations in lattice QCD~\cite{Karsch:2001cy} indicate that the transition should occur at a critical temperature $T_c \simeq$~175~MeV, corresponding to an energy density $\varepsilon_c \simeq$~1~GeV/fm$^3$.

If the QGP is formed in heavy-ion collisions, it is expected to screen the confining potential of heavy quark-antiquark pairs~\cite{Matsui:1986dk}, leading to the melting of charmonium and bottomonium states: \JPsi, $\psi(2S)$, $\chi_c$, $\Upsilon(1S)$, $\Upsilon(2S)$, $\Upsilon(3S)$, and $\chi_b$. The melting temperature depends on the binding energy of the quarkonium state. The ground states, \JPsi and $\Upsilon(1S)$, are expected to dissolve at significantly higher temperatures than the more loosely bound excited states. Quenched lattice QCD calculations~\cite{Satz:2005hx,Wong:2004zr} originally predicted that the $\Upsilon$ states melt at 1.2~$T_c$ ($3S$), 1.6~$T_c$ ($2S$), and above 4~$T_c$ ($1S$), while modern spectral-function approaches with complex potentials~\cite{Miao:2010tk} favour somewhat lower dissolution temperatures. This sequential melting pattern is generally considered a ``smoking-gun'' signature of the QCD deconfinement transition. However, a large fraction of the observed $1S$ yield in elementary collisions is caused by feed-down contributions from decays of heavier states (around 50\% for the $\Upsilon(1S)$~\cite{Affolder:1999wm}). Therefore the melting of the excited states is expected to result in a significant suppression of the observed $1S$ yields, even if the medium is not hot enough to directly dissolve the ground states.

Observations of \JPsi and $\psi(2S)$ suppression between proton-nucleus and heavy-ion collisions were reported by the NA38~\cite{Baglin:1994ui}, NA50~\cite{Alessandro:2004ap, Alessandro:2006ju} and NA60~\cite{Arnaldi:2007zz} fixed-target experiments at the Super Proton Synchrotron (SPS), respectively in SU, PbPb and InIn collisions, at centre-of-mass energies per nucleon pair ($\sqrt{s_{\rm NN}}$) of about 20~GeV. The PHENIX experiment, at the Relativistic Heavy Ion Collider (RHIC), extended the \JPsi suppression measurements to AuAu collisions at $\sqrt{s_{\rm NN}}=200$~GeV~\cite{Adare:2006ns}. At RHIC, bottomonia production becomes measurable~\cite{Abelev:2010am}, though with limited integrated luminosities. PHENIX observed that the dimuon yield in the $\Upsilon$ mass region for minimum bias AuAu collisions is less than 64\%, at the 90\% confidence level, of the value expected by extrapolating the pp yields~\cite{Atomssa:2009ek}.

A new era of detailed studies of the bottomonium family in heavy-ion collisions has started at the Large Hadron Collider (LHC). The measurement reported in this Letter is performed with the data recorded by the Compact Muon Solenoid (CMS) experiment during the first PbPb LHC run, at the end of 2010, and during the \Pp\Pp\ run of March 2011, both at $\sqrt{s_{\rm NN}} = 2.76$~TeV. The integrated luminosity used in this analysis corresponds to 7.28~$\mu$b$^{-1}$ for PbPb and 225~nb$^{-1}$ for \Pp\Pp\ collisions, the latter corresponding approximately to the equivalent nucleon-nucleon luminosity of the PbPb run. The excellent momentum resolution of the CMS detector results in well-resolved $\Upsilon$ peaks in the dimuon mass spectrum. The CMS collaboration has previously studied $\Upsilon$ production in \Pp\Pp\ data at $\sqrt{s}=7$~TeV~\cite{Khachatryan:2010zg}, using techniques to extract the $\Upsilon$ yields that are very similar to the ones used in the study reported in this Letter.

A detailed description of the CMS detector can be found in~\cite{CMS:2008zzk}. Its central feature is a superconducting solenoid of 6~m internal diameter, providing a magnetic field of 3.8~T. Within the field volume are the silicon pixel and strip tracker, the crystal electromagnetic calorimeter, and the brass/scintillator hadron calorimeter. Muons are measured in gas-ionisation detectors embedded in the steel return yoke. In addition, CMS has extensive forward calorimetry, in particular two steel/quartz-fiber \v{C}erenkov hadron forward (HF) calorimeters, which cover the pseudorapidity range $2.9 < |\eta| < 5.2$.

In this analysis, $\Upsilon$ mesons are identified through their dimuon decay. The silicon pixel and strip tracker measures charged-particle trajectories in the  range $|\eta| < 2.5$. The tracker consists of 66M pixel and 10M strip detector channels, providing a vertex resolution of $\sim$\,15~$\mu$m in the transverse plane. Muons are detected in the $|\eta| < 2.4$ range, with detection planes based on three technologies: drift tubes, cathode strip chambers, and resistive plate chambers. Due to the strong magnetic field and the fine granularity of the silicon tracker, the muon transverse momentum measurement ($p_{\rm T}$) based on information from the silicon tracker alone has a resolution between 1 and 2\% for a typical muon in this analysis.

In both the PbPb and \Pp\Pp\ runs, the events are selected by the CMS two-level trigger. At the first, hardware level, two independent muon candidates are required in the muon detectors. No selection is made on momentum or pseudorapidity, but in the \Pp\Pp\ case more stringent quality requirements are imposed for each muon in order to reduce the higher trigger rate. In both cases, the software-based higher-level trigger accepts the lower-level decision without applying further criteria. From reconstructed $JPsi \rightarrow \mu\mu$ decays, the single-muon trigger efficiencies are measured and found to be consistent between the PbPb, $(96.1 \pm 1.0)$\%, and the \Pp\Pp, $(95.5 \pm 0.6)$\%, data sets, for muons with $p_{\rm T} >$~4~GeV/$c$.

In the PbPb data, events are preselected offline if they contain a reconstructed primary vertex made of at least two tracks, and a coincidence in both HF calorimeters of energy deposits in at least three towers of 3~GeV each. These criteria reduce contributions from single-beam interactions (e.g. beam-gas and beam-halo collisions with the beam pipe), ultra-peripheral electromagnetic collisions, and cosmic-ray muons. A small fraction of the most peripheral PbPb collisions is not selected by these requirements, which accept $(97 \pm 3)$\% of the hadronic inelastic cross section~\cite{HIN-10-004}. For the \Pp\Pp\ run, a similar event filter is applied, relaxing the HF coincidence to one tower in each HF, with at least 3~GeV deposited. This filter removes only 1\% of the \Pp\Pp\ events satisfying the dimuon trigger.

The muon offline reconstruction is seeded with $\simeq 99$\% efficiency by tracks in the muon detectors, called standalone muons. These tracks are then matched to tracks reconstructed in the silicon tracker by means of an algorithm optimised for the heavy-ion environment~\cite{Roland:2006kz,D'Enterria:2007xr}. For muons from $\Upsilon$ decays the tracking efficiency is $\simeq$~85\%. This efficiency is lower than in \Pp\Pp, as in PbPb the track reconstruction is seeded by a greater number of pixel hits to reduce the large number of random combinations arising from the high multiplicity of each event. Combined fits of the muon and tracker tracks are used to obtain the results presented in this Letter. The heavy-ion dedicated reconstruction algorithm is applied to the pp data in order to avoid potential biases, arising from different tracking efficiencies of the two reconstruction algorithms, when comparing the two data sets.

Identical very loose selection criteria are applied to the muons in the \Pp\Pp\ and PbPb data. The transverse (longitudinal) distance from the event vertex is required to be less than 3 (15) cm. Tracks are only kept if they have 11 or more hits in the silicon tracker and the $\chi^2$ per degree of freedom of the combined (tracker) track fit is lower than 20 (4). The two muon trajectories are fit with a common vertex constraint, and events are retained if the fit $\chi^2$ probability is larger than 1\%. This removes background arising primarily from displaced heavy quark semileptonic decays. As determined from Monte Carlo simulation of the $\Upsilon(1S)$ signal, these selection criteria are found to reduce the efficiency by 3.9\%, consistent with the signal loss observed in both \Pp\Pp\ and PbPb data. The available event sample limits to 20~GeV/$c$ the dimuon transverse momentum
range probed in this study.

In order to further reduce the background in the $\Upsilon$ mass region, only muons with a transverse momentum ($p_{\rm T}^\mu$) higher than 4~GeV/$c$ are considered, resulting in a $\Upsilon$ acceptance of approximately 25\%  for the $|y^\Upsilon|<2.4$ rapidity range. This requirement improves the significance of the $\Upsilon(1S)$ signal in PbPb data and is applied to both data sets. The acceptance of a $\Upsilon$ state depends on its mass, since the excited states give rise to higher-momenta muons. In consequence, requiring higher $p_{\rm T}^\mu$ increases the acceptance for the excited states relative to the ground state. In the corresponding analysis performed with the higher-statistics (3.1~pb$^{-1}$) 7~TeV data~\cite{Khachatryan:2010zg}, looser criteria were applied ($p_{\rm T}^\mu >$~3.5 GeV/$c$ and $|\eta^\mu| < 1.6$, or $p_{\rm T}^\mu >$2.5 GeV/$c$ and $1.6 < |\eta^\mu| < 2.4$), where $\eta^\mu$ is the muon pseudorapidity. The stricter ($p_{\rm T}^\mu >$~4~GeV/$c$) requirements used here  enhance the  $\Upsilon(2S+3S)/\Upsilon(1S)$ yield ratio by $\simeq 60$\% in the \Pp\Pp\ data at 2.76~TeV. It was checked that, applying the same reconstruction algorithm and the same $p_{\rm T}^\mu$ requirements, the $\Upsilon(2S+3S)/\Upsilon(1S)$ yield ratio is consistent between the 2.76 and 7~TeV \Pp\Pp\ data sets.

\begin{figure}[hbtp]
  \begin{center}
    \includegraphics[width=0.95\linewidth]{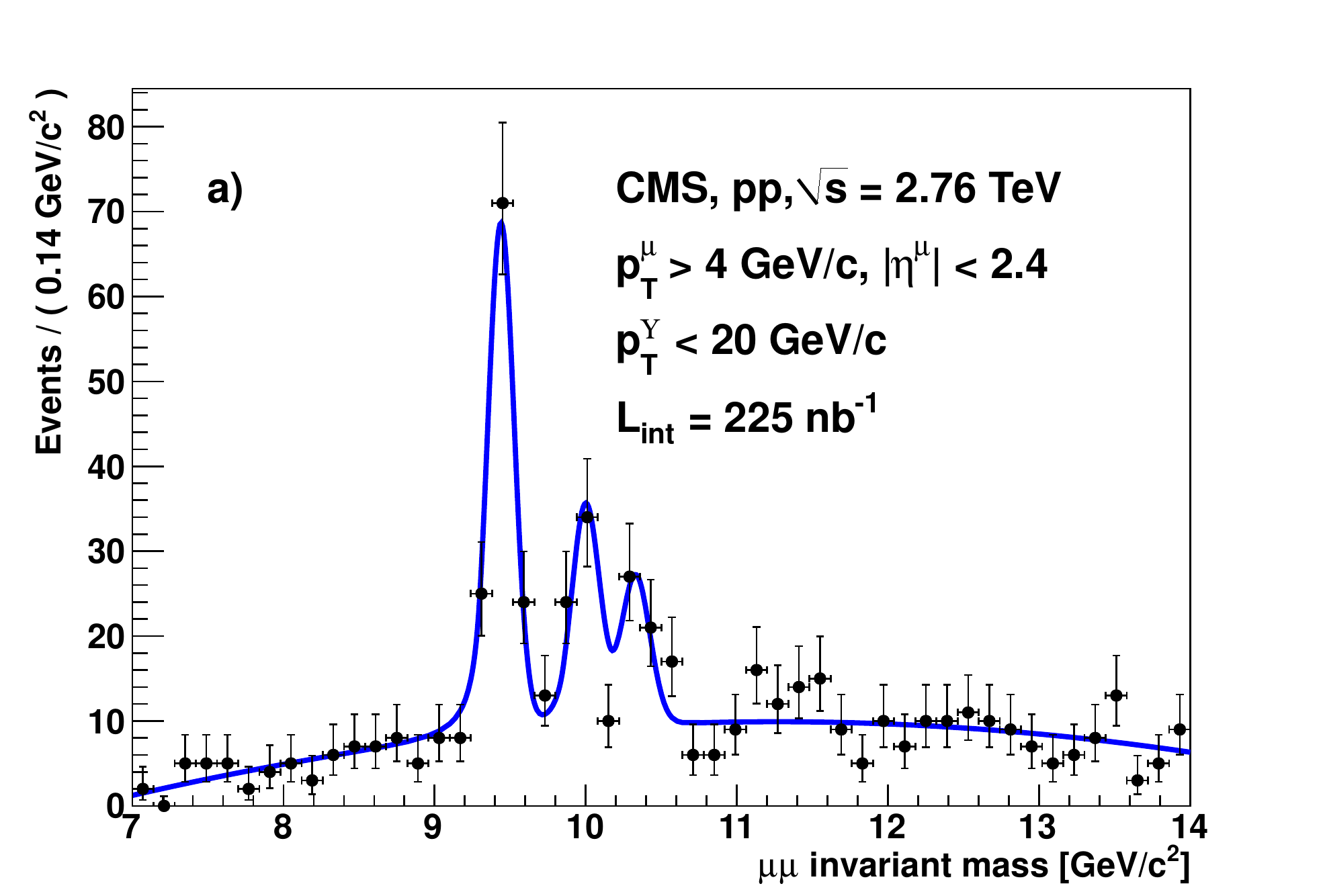} 
    \includegraphics[width=0.95\linewidth]{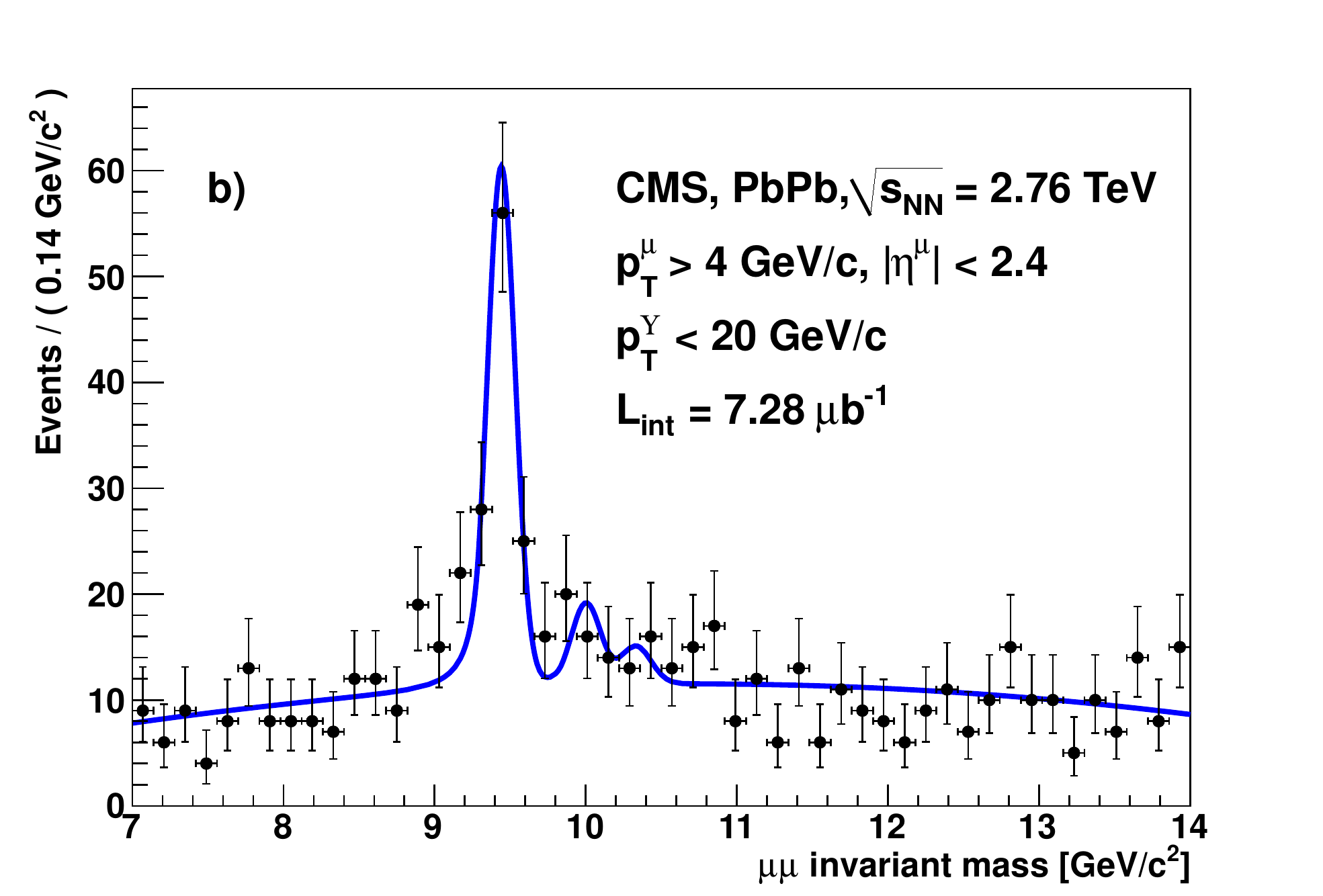} 
    \caption{Dimuon invariant-mass distributions from the \Pp\Pp\ (a) and PbPb (b) data at $\sqrt{s_{\rm NN}} = 2.76$~TeV. The same reconstruction algorithm and analysis criteria are applied to both data sets, including a transverse momentum requirement on single muons of $p_{\rm T}^\mu >$~4~GeV/$c$. The solid lines show the result of the fit described in the text. }
    \label{fig:mass}
  \end{center}
    \vspace{-2ex} 
\end{figure}

The dimuon invariant mass spectra with the selection criteria applied are shown in Fig.~\ref{fig:mass} for the \Pp\Pp\ and PbPb data sets. Within the 7--14~GeV/$c^2$ mass range, there are 561 (628) opposite-sign muon pairs in the \Pp\Pp\ (PbPb) data set. The three Y peaks are clearly observed in the \Pp\Pp\ case, but the $\Upsilon(2S)$ and $\Upsilon(3S)$ are barely visible over the residual background in PbPb collisions.

An extended unbinned maximum likelihood fit to the two invariant mass distributions of Fig.~\ref{fig:mass} is performed to extract the yields, following the method described in~\cite{Khachatryan:2010zg}. The measured mass lineshape of each $\Upsilon$ state is parameterised by a ``Crystal Ball'' (CB) function, i.e. a Gaussian resolution function with the low-side tail replaced by a power law describing final state radiation (FSR). Since the three $\Upsilon$ resonances partially overlap in the measured dimuon mass, they are fit simultaneously. Therefore, the probability distribution function (PDF) describing the signal consists of three CB functions. In addition to the three $\Upsilon(nS)$ yields, the $\Upsilon(1S)$ mass is the only parameter left free, to accommodate a possible bias in the momentum scale calibration. The mass differences between the states are fixed to their world average values~\cite{Nakamura:2010zzi} and the mass resolution is forced to scale with the resonance mass. The $\Upsilon(1S)$ resolution is fixed to the value estimated in the simulation, 92~MeV/$c^2$, which is compatible with the resolution obtained from both the PbPb and \Pp\Pp\ data. The low-side tail parameters are also fixed to the values obtained via simulation. Finally, a second-order polynomial is chosen to describe the background in the 7--14~GeV/$c^2$ mass-fit range.

The quality of the unbinned fit is checked \emph{a posteriori} by comparing the obtained lineshapes to the binned data of Fig.~\ref{fig:mass}. The $\chi^2$ probabilities are 74\% and 77\%, respectively for \Pp\Pp\ and PbPb.

The ratios of the observed yields of the $\Upsilon(2S)$ and $\Upsilon(3S)$ excited states to the $\Upsilon(1S)$ ground state in the \Pp\Pp\ and PbPb data are:
\begin{eqnarray}
  \Upsilon(2S+3S)/\Upsilon(1S)|_{\Pp\Pp} & = & 0.78 _{-0.14}^{+0.16} \pm 0.02,\\
  \Upsilon(2S+3S)/\Upsilon(1S)|_{\rm PbPb} & = & 0.24 _{-0.12}^{+0.13} \pm 0.02,
\end{eqnarray}
where the first uncertainty is statistical and the second is systematic.

The systematic uncertainties are computed by varying the lineshape in the following ways: (1) the CB-tail parameters are varied randomly according to their covariance matrix and within conservative values covering imperfect knowledge of the amount of detector material and FSR in the underlying process; (2) the resolution is varied by $\pm 5$~MeV/$c^2$, which is a conservative variation given the current understanding of the detector performance and reasonable changes that can be anticipated in the $\Upsilon$-resonance kinematics between \Pp\Pp\ and PbPb data; (3) the background shape is changed from quadratic to linear while the mass range of the fit is varied from 6--15 to 8--12~GeV/$c^2$; the observed root-mean-square of the results is taken as the systematic uncertainty. The quadrature sum of these three systematic uncertainties gives a relative uncertainty on the ratio of 10\% (3\%) on the PbPb (\Pp\Pp) data.

The ratio of the $\Upsilon(2S+3S)/\Upsilon(1S)$ ratios in PbPb and \Pp\Pp\ benefits from an almost complete cancellation of possible acceptance and/or efficiency differences among the reconstructed resonances. A simultaneous fit to the \Pp\Pp\ and PbPb mass spectra gives the double ratio
\begin{equation}
  \frac{\Upsilon(2S+3S)/\Upsilon(1S)|_{\rm PbPb}}{\Upsilon(2S+3S)/\Upsilon(1S)|_{\Pp\Pp}}
 = 0.31 _{-0.15}^{+0.19} \; ({\rm stat.}) \pm 0.03 \; ({\rm syst.}),
\label{eq:double}
\end{equation}
where the systematic uncertainty (9\%) arises from varying the lineshape as described above in the simultaneous fit, thus taking into account partial cancellations of systematic effects.

The single muon lower momentum requirement is \emph{a posteriori} varied from 3 to 5~GeV/$c$ in steps of 500~MeV/$c$, and it is found that $p_{\rm T}$ requirements other than 4~GeV/$c$ provide lower double ratios. Fitting the \Pp\Pp\ and PbPb spectra with free and independent mass resolution parameters
leads to an increase of the double ratio by 15\%.

To evaluate possible imperfect cancellations of acceptance and efficiency effects in the double ratio, a full \GEANTfour~\cite{Agostinelli:2002hh} detector simulation is performed. The effect of the higher PbPb underlying event activity is considered by embedding, at the level of detector signals, $\Upsilon(1S)$ and $\Upsilon(2S)$ decays simulated by \PYTHIA 6.424~\cite{Ref:Pythia} in PbPb events simulated with \textsc{hydjet}~\cite{Lokhtin:2005px}. Track characteristics, such as the number of hits and the $\chi^2$ of the track fit, have similar distributions in data and simulation. As mentioned above, the trigger efficiency is evaluated with data, by using single-muon-triggered data events, and reconstructing \JPsi signal with and without the dimuon trigger requirement. The same exercise is carried out with the simulation and it agrees with the efficiency measured in data at the 2\% level. The track efficiency in the silicon detector is measured with standalone muons, applying all selection criteria. The efficiencies in data and simulation agree within the 4\% statistical uncertainty of the efficiency determined from data.

The difference in reconstruction and selection efficiencies between the $\Upsilon$ states is less than 5\% and the variation with charged particle multiplicity is less than 10\% from \Pp\Pp\ to central PbPb collisions, producing a maximum change of 0.5\% on the double ratio. The good agreement between single-muon trigger efficiencies extracted from data for the \Pp\Pp\ and PbPb trigger requirements, applied to the $\Upsilon(1S)$ and $\Upsilon(2S)$ trigger efficiencies derived from simulation, leads to a negligible effect on the double ratio. The single-muon trigger efficiencies extracted from data agree within 1.5\% for the \Pp\Pp\ and PbPb trigger requirements, and the $\Upsilon(1S)$ and $\Upsilon(2S)$ trigger efficiencies agree within 3\%, according to simulation: the potential trigger bias on the double ratio is negligible. The magnitudes of the statistical and systematic uncertainties on the double ratio, respectively 55\% and 9\%, are significantly larger than the systematic uncertainties associated with possible imperfect cancellation of acceptance and efficiency effects. Therefore no additional uncertainty from these sources is applied.

Finally, using an ensemble of one million pseudo-experiments, generated with the signal lineshape obtained from the \Pp\Pp\ data (Fig.~\ref{fig:mass}a), the background lineshapes from both data sets, and a double ratio (Eq.~\ref{eq:double}) equal to unity within uncertainties, the probability of finding the measured value of 0.31 or below is estimated to be 0.9\%. In other words, in the absence of a suppression due to physics mechanisms, the probability of a downward departure of the ratio from unity of this significance or greater is 0.9\%, i.e. that corresponding to 2.4 sigma in a one-tailed integral of a Gaussian distribution.

Other studies from the CMS experiment show that the $\Upsilon(1S)$ itself is suppressed by about 40\%~\cite{HIN-10-006} in minimum bias PbPb collisions at $\sqrt{s_{\rm NN}} = 2.76$~TeV. Since a large fraction of the $\Upsilon(1S)$ yield arises from decays of heavier bottomonium states~\cite{Affolder:1999wm}, this $\Upsilon(1S)$ suppression could be indirectly caused by the suppression of the excited states reported in this Letter.

Production yields of quarkonium states can also be modified, from \Pp\Pp\ to PbPb collisions, in the absence of QGP formation, by cold nuclear matter effects~\cite{Vogt:2010aa}. However, such effects should have a small impact on the $\Upsilon$ double ratio reported here. The nuclear modifications of the parton distribution functions (shadowing) should have an equivalent effect on the three $\Upsilon$ states, because their production involves very similar partons, cancelling in the ratio, at least to first order. The same should happen to any other initial-state nuclear effect. In principle, the larger and more loosely bound excited quarkonium states are more likely to be broken up by final-state interactions while traversing the nuclear matter, something extensively studied in the context of charmonium suppression at lower energies~\cite{Lourenco:2008sk}. This ``nuclear absorption'' becomes weaker with increasing energy, and should be negligible at the LHC. At RHIC energies, the STAR experiment~\cite{Reed:2011zz} has reported a $\Upsilon(1S+2S+3S)$ yield in dAu collisions of $0.78 \pm 0.28 \pm 0.20$ times the yield expected by scaling \Pp\Pp\ collisions, compatible with the absence of absorption. Furthermore, the double ratio presented here would only be sensitive to a \emph{difference} between the nuclear dependencies of the three states and already at much lower energies the Fermilab E772 experiment observed~\cite{Alde:1991sw}, in proton-nucleus collisions, no such difference, within uncertainties, between the $\Upsilon(1S)$ and the sum $\Upsilon(2S+3S)$.

Future high-statistics heavy-ion and proton-nucleus runs at the LHC will provide further quarkonia measurements, which should help disentangle nuclear from medium effects and aid the interpretation of the result reported in this Letter.

In summary, a comparison of the relative yields of $\Upsilon$ resonances has been performed in PbPb and \Pp\Pp\ collisions at the same centre-of-mass energy per nucleon pair of 2.76 TeV. The double ratio of the $\Upsilon(2S)$ and $\Upsilon(3S)$ excited states to the $\Upsilon(1S)$ ground state in PbPb and \Pp\Pp\ collisions, $[\Upsilon(2S+3S)/\Upsilon(1S)]_{\rm PbPb}/[\Upsilon(2S+3S)/\Upsilon(1S)]_{\Pp\Pp}$, is found to be 0.31~$_{-0.15}^{+0.19}$~(stat.)~$\pm$~0.03~(syst.), for muons of $p_{\rm T}>4$~GeV/$c$ and $|\eta|<2.4$. The probability to obtain the measured value, or lower, if the true double ratio is unity, has been calculated to be less than 1\%.

\section*{Acknowledgments}

We wish to congratulate our colleagues in the CERN accelerator departments for the excellent performance of the LHC machine. We thank the technical and administrative staff at CERN and other CMS institutes, and acknowledge support from: FMSR (Austria); FNRS and FWO (Belgium); CNPq, CAPES, FAPERJ, and FAPESP (Brazil); MES (Bulgaria); CERN; CAS, MoST, and NSFC (China); COLCIENCIAS (Colombia); MSES (Croatia); RPF (Cyprus); Academy of Sciences and NICPB (Estonia); Academy of Finland, ME, and HIP (Finland); CEA and CNRS/IN2P3 (France); BMBF, DFG, and HGF (Germany); GSRT (Greece); OTKA an NKTH (Hungary); DAE and DST (India); IPM (Iran); SFI (Ireland); INFN (Italy); NRF and WCU (Korea); LAS (Lithuania); CINVESTAV, CONACYT, SEP, and UASLP-FAI (Mexico); PAEC (Pakistan); SCSR (Poland); FCT (Portugal); JINR (Armenia, Belarus, Georgia, Ukraine, Uzbekistan); MST and MAE (Russia); MSTD (Serbia); MICINN and CPAN (Spain); Swiss Funding Agencies (Switzerland); NSC (Taipei); TUBITAK and TAEK (Turkey); STFC (United Kingdom); DOE and NSF (USA). Individuals have received support from the Marie-Curie programme and the European Research Council (European Union); the Leventis Foundation; the A. P. Sloan Foundation; the Alexander von Humboldt Foundation; the Associazione per lo Sviluppo Scientifico e Tecnologico del Piemonte (Italy); the Belgian Federal Science Policy Office; the Fonds pour la Formation \`a la Recherche dans l'industrie et dans l'Agriculture (FRIA-Belgium); and the Agentschap voor Innovatie door Wetenschap en Technologie (IWT-Belgium).

\bibliography{auto_generated}   

\providecommand{\href}[2]{#2}\begingroup\raggedright\begin{thebibliography}{10}%
\makeatletter
\providecommand{\hrefCMSnoop }[0]{\@secondoftwo}%
\makeatother

\bibitem{Karsch:2001cy}
\hrefCMSnoop {} {F.~Karsch, ``Lattice QCD at high temperature and density'',}
  \textit{ Lect. Notes Phys.} \textbf{ 583} (2002) 209,
\href{http://www.arXiv.org/abs/hep-lat/0106019}{\texttt{
  arXiv:hep-lat/0106019}}.

\bibitem{Matsui:1986dk}
\hrefCMSnoop {} {T.~Matsui and H.~Satz, ``{$J/\psi$ Suppression by Quark-Gluon
  Plasma Formation}'',} \textit{ Phys. Lett.} \textbf{ B178} (1986) 416.
\href{http://dx.doi.org/10.1016/0370-2693(86)91404-8}{\texttt{
  doi:10.1016/0370-2693(86)91404-8}}.

\bibitem{Satz:2005hx}
\hrefCMSnoop {} {H.~Satz, ``{Colour deconfinement and quarkonium binding}'',}
  \textit{ J. Phys.} \textbf{ G32} (2006) R25,
  \href{http://www.arXiv.org/abs/hep-ph/0512217}{\texttt{
  arXiv:hep-ph/0512217}}.
\href{http://dx.doi.org/10.1088/0954-3899/32/3/R01}{\texttt{
  doi:10.1088/0954-3899/32/3/R01}}.

\bibitem{Wong:2004zr}
\hrefCMSnoop {} {C.-Y. Wong, ``{Heavy quarkonia in quark gluon plasma}'',}
  \textit{ Phys. Rev.} \textbf{ C72} (2005) 034906,
  \href{http://www.arXiv.org/abs/hep-ph/0408020}{\texttt{
  arXiv:hep-ph/0408020}}.
\href{http://dx.doi.org/10.1103/PhysRevC.72.034906}{\texttt{
  doi:10.1103/PhysRevC.72.034906}}.

\bibitem{Miao:2010tk}
\hrefCMSnoop {} {C.~Miao, A.~Mocsy, and P.~Petreczky, ``Quarkonium spectral
  functions with complex potential'',} \textit{ Nucl. Phys.} \textbf{ A855}
  (2011) 125, \href{http://www.arXiv.org/abs/1012.4433}{\texttt{
  arXiv:1012.4433}}.
\href{http://dx.doi.org/10.1016/j.nuclphysa.2011.02.028}{\texttt{
  doi:10.1016/j.nuclphysa.2011.02.028}}.

\bibitem{Affolder:1999wm}
\hrefCMSnoop {} {{ CDF} Collaboration, ``{Production of $\Upsilon(1S)$ mesons
  from $\chi_b$ decays in $p\bar{p}$ collisions at $\sqrt{s} = 1.8$ TeV}'',}
  \textit{ Phys. Rev. Lett.} \textbf{ 84} (2000) 2094,
  \href{http://www.arXiv.org/abs/hep-ex/9910025}{\texttt{
  arXiv:hep-ex/9910025}}.
\href{http://dx.doi.org/10.1103/PhysRevLett.84.2094}{\texttt{
  doi:10.1103/PhysRevLett.84.2094}}.

\bibitem{Baglin:1994ui}
\hrefCMSnoop {} {{ NA38} Collaboration, ``{$\psi'$ and $J/\psi$ production in p
  W, p U and S U interactions at 200-GeV/nucleon}'',} \textit{ Phys. Lett.}
  \textbf{ B345} (1995) 617.
\href{http://dx.doi.org/10.1016/0370-2693(94)01614-I}{\texttt{
  doi:10.1016/0370-2693(94)01614-I}}.

\bibitem{Alessandro:2004ap}
\hrefCMSnoop {} {{ NA50} Collaboration, ``{A new measurement of $J/\psi$
  suppression in Pb-Pb collisions at 158~GeV per nucleon}'',} \textit{ Eur.
  Phys. J.} \textbf{ C39} (2005) 335,
  \href{http://www.arXiv.org/abs/hep-ex/0412036}{\texttt{
  arXiv:hep-ex/0412036}}.
\href{http://dx.doi.org/10.1140/epjc/s2004-02107-9}{\texttt{
  doi:10.1140/epjc/s2004-02107-9}}.

\bibitem{Alessandro:2006ju}
\hrefCMSnoop {} {{ NA50} Collaboration, ``{$\psi'$ production in Pb-Pb
  collisions at 158 GeV/nucleon}'',} \textit{ Eur. Phys. J.} \textbf{ C49}
  (2007) 559, \href{http://www.arXiv.org/abs/nucl-ex/0612013}{\texttt{
  arXiv:nucl-ex/0612013}}.
\href{http://dx.doi.org/10.1140/epjc/s10052-006-0153-y}{\texttt{
  doi:10.1140/epjc/s10052-006-0153-y}}.

\bibitem{Arnaldi:2007zz}
\hrefCMSnoop {} {{ NA60} Collaboration, ``{$J/\psi$ production in indium-indium
  collisions at 158-GeV/nucleon}'',} \textit{ Phys. Rev. Lett.} \textbf{ 99}
  (2007) 132302.
\href{http://dx.doi.org/10.1103/PhysRevLett.99.132302}{\texttt{
  doi:10.1103/PhysRevLett.99.132302}}.

\bibitem{Adare:2006ns}
\hrefCMSnoop {} {{ PHENIX} Collaboration, ``$J/\psi$ production vs centrality,
  transverse momentum, and rapidity in Au + Au collisions at $\sqrt{s_{NN}} =
  200$~GeV'',} \textit{ Phys. Rev. Lett.} \textbf{ 98} (2007) 232201,
  \href{http://www.arXiv.org/abs/nucl-ex/0611020}{\texttt{
  arXiv:nucl-ex/0611020}}.
\href{http://dx.doi.org/10.1103/PhysRevLett.98.232301}{\texttt{
  doi:10.1103/PhysRevLett.98.232301}}.

\bibitem{Abelev:2010am}
\hrefCMSnoop {} {{ STAR} Collaboration, ``{Upsilon cross section in p+p
  collisions at sqrt(s) = 200 GeV}'',} \textit{ Phys. Rev.} \textbf{ D82}
  (2010) 012004, \href{http://www.arXiv.org/abs/1001.2745}{\texttt{
  arXiv:1001.2745}}.
\href{http://dx.doi.org/10.1103/PhysRevD.82.012004}{\texttt{
  doi:10.1103/PhysRevD.82.012004}}.

\bibitem{Atomssa:2009ek}
\hrefCMSnoop {} {{ PHENIX} Collaboration, ``{$J/\psi$ Elliptic Flow, High
  $p_{\rm T}$ Suppression and Upsilon Measurements in A+A Collisions by the
  PHENIX Experiment}'',} \textit{ Nucl. Phys.} \textbf{ A830} (2009) 331c,
  \href{http://www.arXiv.org/abs/0907.4787}{\texttt{ arXiv:0907.4787}}.
\href{http://dx.doi.org/10.1016/j.nuclphysa.2009.10.099}{\texttt{
  doi:10.1016/j.nuclphysa.2009.10.099}}.

\bibitem{Khachatryan:2010zg}
\hrefCMSnoop {} {{ CMS} Collaboration, ``Upsilon production cross section in pp
  collisions at $\sqrt{s}=7$~{T}eV'',} (2011).
  \href{http://www.arXiv.org/abs/1012.5545}{\texttt{ arXiv:1012.5545}}.
  Accepted for publication by \textit{Phys. Rev. D}.


\bibitem{CMS:2008zzk}
\hrefCMSnoop {} {{ CMS} Collaboration, ``{The CMS experiment at the CERN
  LHC}'',} \textit{ JINST} \textbf{ 3} (2008) S08004.
\href{http://dx.doi.org/10.1088/1748-0221/3/08/S08004}{\texttt{
  doi:10.1088/1748-0221/3/08/S08004}}.

\bibitem{HIN-10-004}
\hrefCMSnoop {} {{ CMS} Collaboration, ``Observation and studies of jet
  quenching using dijet momentum imbalance in {PbPb} collisions at
  $\sqrt{s_{NN}}=2.76$~{TeV} with the {CMS} detector'',} (2011).
  \href{http://www.arXiv.org/abs/1102.1957}{\texttt{ arXiv:1102.1957}}.
  Submitted to \textit{Phys. Rev.~C}.

\bibitem{Roland:2006kz}
\hrefCMSnoop {} {{ CMS} Collaboration, ``{Track reconstruction in heavy ion
  events using the CMS tracker}'',} \textit{ Nucl. Instrum. Meth.} \textbf{
  A566} (2006) 123.
\href{http://dx.doi.org/10.1016/j.nima.2006.05.023}{\texttt{
  doi:10.1016/j.nima.2006.05.023}}.

\bibitem{D'Enterria:2007xr}
\hrefCMSnoop {} {{ CMS} Collaboration, ``{CMS physics technical design report:
  Addendum on high density QCD with heavy ions}'',} \textit{ J. Phys.} \textbf{
  G34} (2007) 2307.
  \href{http://dx.doi.org/10.1088/0954-3899/34/11/008}{\texttt{
  doi:10.1088/0954-3899/34/11/008}}.

\bibitem{Nakamura:2010zzi}
\hrefCMSnoop {} {{ Particle Data Group} Collaboration, ``{Review of particle
  physics}'',} \textit{ J. Phys.} \textbf{ G37} (2010) 075021.
\href{http://dx.doi.org/10.1088/0954-3899/37/7A/075021}{\texttt{
  doi:10.1088/0954-3899/37/7A/075021}}.

\bibitem{Agostinelli:2002hh}
\hrefCMSnoop {} {{ GEANT4} Collaboration, ``{GEANT4: A simulation toolkit}'',}
  \textit{ Nucl. Instrum. Meth.} \textbf{ A506} (2003) 250.
\href{http://dx.doi.org/10.1016/S0168-9002(03)01368-8}{\texttt{
  doi:10.1016/S0168-9002(03)01368-8}}.

\bibitem{Ref:Pythia}
\hrefCMSnoop {} {T.~Sj{\"o}strand, S.~Mrenna, and P.~Skands, ``PYTHIA 6.4
  physics and manual'',} \textit{ JHEP} \textbf{ 05} (2006) 026,
  \href{http://www.arXiv.org/abs/hep-ph/0603175}{\texttt{
  arXiv:hep-ph/0603175}}.
  \href{http://dx.doi.org/10.1088/1126-6708/2006/05/026}{\texttt{
  doi:10.1088/1126-6708/2006/05/026}}.

\bibitem{Lokhtin:2005px}
\hrefCMSnoop {} {I.~P. Lokhtin and A.~M. Snigirev, ``{A model of jet quenching
  in ultrarelativistic heavy ion collisions and high-p(T) hadron spectra at
  RHIC}'',} \textit{ Eur. Phys. J.} \textbf{ C45} (2006) 211--217,
  \href{http://www.arXiv.org/abs/hep-ph/0506189}{\texttt{
  arXiv:hep-ph/0506189}}.
\href{http://dx.doi.org/10.1140/epjc/s2005-02426-3}{\texttt{
  doi:10.1140/epjc/s2005-02426-3}}.

\bibitem{HIN-10-006}
\hrefCMSnoop {} {{ CMS} Collaboration, ``Quarkonium production in {PbPb}
  collisions at $\sqrt{s_{\rm NN}}$=2.76~{TeV}'',} (2011).
  {CMS-PAS-HIN-10-006}. To be released.

\bibitem{Vogt:2010aa}
\hrefCMSnoop {} {R.~Vogt, ``{Cold Nuclear Matter Effects on J/psi and Upsilon
  Production at the LHC}'',} \textit{ Phys. Rev.} \textbf{ C81} (2010) 044903,
  \href{http://www.arXiv.org/abs/1003.3497}{\texttt{ arXiv:1003.3497}}.
\href{http://dx.doi.org/10.1103/PhysRevC.81.044903}{\texttt{
  doi:10.1103/PhysRevC.81.044903}}.

\bibitem{Lourenco:2008sk}
\hrefCMSnoop {} {C.~Lourenco, R.~Vogt, and H.~K. Woehri, ``{Energy dependence
  of J/psi absorption in proton-nucleus collisions}'',} \textit{ JHEP} \textbf{
  02} (2009) 014, \href{http://www.arXiv.org/abs/0901.3054}{\texttt{
  arXiv:0901.3054}}.
\href{http://dx.doi.org/10.1088/1126-6708/2009/02/014}{\texttt{
  doi:10.1088/1126-6708/2009/02/014}}.

\bibitem{Reed:2011zz}
\hrefCMSnoop {} {{ STAR} Collaboration, ``{Upsilon production in p + p, d + Au,
  Au + Au collisions at $\sqrt{s_{NN}} = 200$~GeV in STAR}'',} \textit{ J.
  Phys. Conf. Ser.} \textbf{ 270} (2011) 012026.
\href{http://dx.doi.org/10.1088/1742-6596/270/1/012026}{\texttt{
  doi:10.1088/1742-6596/270/1/012026}}.

\bibitem{Alde:1991sw}
\hrefCMSnoop {} {D.~M. Alde {et~al.}, ``{Nuclear dependence of the production
  of Upsilon resonances at 800-GeV}'',} \textit{ Phys. Rev. Lett.} \textbf{ 66}
  (1991) 2285.
\href{http://dx.doi.org/10.1103/PhysRevLett.66.2285}{\texttt{
  doi:10.1103/PhysRevLett.66.2285}}.

\end{thebibliography}\endgroup

\cleardoublepage\appendix\section{The CMS Collaboration \label{app:collab}}\begin{sloppypar}\hyphenpenalty=5000\widowpenalty=500\clubpenalty=5000\textbf{Yerevan Physics Institute,  Yerevan,  Armenia}\\*[0pt]
S.~Chatrchyan, V.~Khachatryan, A.M.~Sirunyan, A.~Tumasyan
\vskip\cmsinstskip
\textbf{Institut f\"{u}r Hochenergiephysik der OeAW,  Wien,  Austria}\\*[0pt]
W.~Adam, T.~Bergauer, M.~Dragicevic, J.~Er\"{o}, C.~Fabjan, M.~Friedl, R.~Fr\"{u}hwirth, V.M.~Ghete, J.~Hammer\cmsAuthorMark{1}, S.~H\"{a}nsel, M.~Hoch, N.~H\"{o}rmann, J.~Hrubec, M.~Jeitler, W.~Kiesenhofer, M.~Krammer, D.~Liko, I.~Mikulec, M.~Pernicka, B.~Rahbaran, H.~Rohringer, R.~Sch\"{o}fbeck, J.~Strauss, A.~Taurok, F.~Teischinger, P.~Wagner, W.~Waltenberger, G.~Walzel, E.~Widl, C.-E.~Wulz
% ...
\vskip\cmsinstskip
% ...
1:~~Also at CERN, European Organization for Nuclear Research, Geneva, Switzerland\\
% ...

\end{sloppypar}
\end{document}